\documentclass[twocolumn,english,aps,pra,showpacs]{revtex4}
\usepackage[T1]{fontenc}
\usepackage[latin1]{inputenc}
\usepackage{amsmath}
\usepackage{amssymb}
\usepackage{graphicx}
\usepackage{esint}

\makeatletter
\@ifundefined{textcolor}{}
{%
 \definecolor{BLACK}{gray}{0}
 \definecolor{WHITE}{gray}{1}
 \definecolor{RED}{rgb}{1,0,0}
 \definecolor{GREEN}{rgb}{0,1,0}
 \definecolor{BLUE}{rgb}{0,0,1}
 \definecolor{CYAN}{cmyk}{1,0,0,0}
 \definecolor{MAGENTA}{cmyk}{0,1,0,0}
 \definecolor{YELLOW}{cmyk}{0,0,1,0}
}

\usepackage{babel}

\makeatother

\usepackage{babel}
\begin{document}

\title{Radio-frequency spectroscopy of weakly bound molecules in spin-orbit
coupled atomic Fermi gases}

\author{Hui Hu$^{1}$, Han Pu$^{2}$, Jing Zhang$^{3}$, Shi-Guo Peng$^{4}$,
and Xia-Ji Liu$^{1}$}

\email{xiajiliu@swin.edu.au}

\selectlanguage{english}%

\affiliation{$^{1}$ARC Centre of Excellence for Quantum-Atom Optics and Centre
for Atom Optics and Ultrafast Spectroscopy, Swinburne University of
Technology, Melbourne 3122, Australia\\
 $^{2}$Department of Physics and Astronomy and Rice Quantum Institute,
Rice University, Houston, TX 77251, USA\\
 $^{3}$State Key Laboratory of Quantum Optics and Quantum Optics
Devices, Institute of Opto-Electronics, Shanxi University, Taiyuan
030006, P. R. China\\
 $^{4}$State Key Laboratory of Magnetic Resonance and Atomic and
Molecular Physics, Wuhan Institute of Physics and Mathematics, Chinese
Academy of Sciences, Wuhan 430071, P. R. China }

\date{\today}
\begin{abstract}
We investigate theoretically radio-frequency spectroscopy of weakly
bound molecules in an ultracold spin-orbit-coupled atomic Fermi gas.
We consider two cases with either equal Rashba and Dresselhaus coupling
or pure Rashba coupling. The former system has been realized very
recently at Shanxi University {[}Wang \textit{et al.}, arXiv:1204.1887{]}
and MIT {[}Cheuk \textit{et al.}, arXiv:1205.3483{]}. We predict realistic
radio-frequency signals for revealing the unique properties of anisotropic
molecules formed by spin-orbit coupling. 
\end{abstract}


\maketitle

\section{Introduction}

The coupling between the spin of electrons to their orbital motion,
the so-called spin-orbit coupling, lies at the heart of a variety
of intriguing phenomena in diverse fields of physics. It is responsible
for the well-known fine structure of atomic spectra in atomic physics,
as well as the recently discovered topological state of matter in
solid-state physics, such as topological insulators and superconductors
\cite{Qi2010,Hasan2010}. For electrons, spin-orbit coupling is a
relativistic effect and in general not strong. Most recently, in a
milestone experiment at the National Institute of Standards and Technology
(NIST), synthetic spin-orbit coupling was created and detected in
an atomic Bose-Einstein condensate (BEC) of $^{87}$Rb atoms \cite{Lin2011}.
Using the same experimental technique, non-interacting spin-orbit-coupled
Fermi gases of $^{40}$K atoms and $^{6}$Li atoms have also been
realized, respectively, at Shanxi University \cite{exptShanXi} and
at Massachusetts Institute of Technology (MIT) \cite{exptMIT}. These
experiments have paved an entirely new way to investigate the celebrated
effects of spin-orbit coupling.

Owing to the high-controllability of ultracold atoms in atomic species,
interactions, confining geometry and purity, the advantage of using
synthetic spin-orbit coupling is apparent: (i) The strength of spin-orbit
coupling between ultracold atoms can be made very strong, much stronger
than that in solids; (ii) New bosonic topological states that have
no analogy in solid-state systems may be created; (iii) Ultracold
atoms are able to realize topological superfluids that are yet to
be observed in the solid state; (iv) Strongly correlated topological
states can be readily realized, whose understanding remains a grand
challenge. At the moment, there has been a flood of theoretical work
on synthetic spin-orbit coupling in BECs \cite{Stanescu2008,Larson2009,Wang2010,Wu2011,Ho2011,Xu2011,Sinha2011,Hu2012a,Hu2012b,Ram2012,Barnett2012,Deng2012,Zhu2011a}
and atomic Fermi gases \cite{Vyasanakere2011,Iskin2011,Zhu2011b,Yu2011,Hu2011,Jiang2011,Gong2011,Han2012,Liu2012a,Cui2012,Liu2012b,He2012,Yi2012},
addressing particularly new exotic superfluid phases arising from
spin-orbit coupling \cite{Wang2010,Wu2011,Hu2012a,Hu2011}.

In this paper, we investigate theoretically momentum-resolved radio-frequency
(rf) spectroscopy of an {\em interacting} two-component atomic
Fermi gas with spin-orbit coupling. The interatomic interactions can
be easily manipulated using a Feshbach resonance in $^{40}$K or $^{6}$Li
atoms \cite{Chin2010}. It is known that weakly bound molecules with
anisotropic mass and anisotropic wave function (in momentum space)
may be formed due to spin-orbit coupling \cite{Vyasanakere2011,Yu2011,Hu2011}.
Here, we aim to predict observable rf signals of these anisotropic
molecules. Our calculation is based on the Fermi's golden rule for
the bound-free transition of a stationary molecule \cite{Chin2005}.
We consider two kinds of spin-orbit coupling: (1) the equal Rashba
and Dresselhaus coupling, $\lambda k_{x}\sigma_{y}$, which has been
realized experimentally at Shanxi University \cite{exptShanXi} and
MIT \cite{exptMIT}, and (2) the pure Rashba coupling, $\lambda(k_{y}\sigma_{x}-k_{x}\sigma_{y})$,which
is yet to be realized. Here $\sigma_{x}$ and $\sigma_{y}$ are the
Pauli matrices, $k_{x}$ and $k_{y}$ are momenta and $\lambda$ is
the strength of spin-orbit coupling. The latter case with pure Rashba
spin-orbit coupling is of particular theoretical interest, since molecules
induced by spin-orbit coupling exist even for a negative \textit{s}-wave
scattering length above Feshbach resonances \cite{Vyasanakere2011,Yu2011,Hu2011}.

The paper is structured as follows. In the next section, we discuss
briefly the rf spectroscopy and the Fermi's golden rule for the calculation
of rf transfer strength. In Sec. III, we consider the experimental
case of equal Rashba and Dresselhaus spin-orbit coupling. We introduce
first the model Hamiltonian and the single-particle and two-particle
wave functions. We then derive an analytic expression for momentum-resolved
rf transfer strength following the Fermi's golden rule. It can be
written explicitly in terms of the two-body wave function. We discuss
in detail the distinct features of momentum-resolved rf spectroscopy
in the presence of spin-orbit coupling, with the use of realistic
experimental parameters. In Sec. IV, we consider an atomic Fermi gas
with pure Rashba spin-orbit coupling, a system anticipated to be realized
in the near future. Finally, in Sec. V, we conclude and make some
final remarks.

\section{Radio-frequency spectroscopy and the Fermi's golden rule}

Radio-frequency spectroscopy, including momentum-resolved rf-spectroscopy,
is a powerful tool to characterize interacting many-body systems.
It has been widely used to study fermionic pairing in a two-component
atomic Fermi gas near Feshbach resonances when it crosses from a Bardeen-Cooper-Schrieffer
(BCS) superfluid of weakly interacting Cooper pairs into a BEC of
tightly bound molecules \cite{Chin2004,Schunck2008,Zhang2012}. Most
recently, it has also been used to detect new quasiparticles known
as repulsive polarons \cite{Kohstall2012,Koschorreck2012}, which
occur when ``impurity'' fermionic particles interact repulsively
with a fermionic environment.

The underlying mechanics of rf-spectroscopy is simple. For an atomic
Fermi gas with two hyperfine states, denoted as $\left|1\right\rangle =\left|\uparrow\right\rangle $
and $\left|2\right\rangle =\left|\downarrow\right\rangle $, the rf
field drives transitions between one of the hyperfine states (i.e.,
$\left|\downarrow\right\rangle $) and an empty hyperfine state $\left|3\right\rangle $
which lies above it by an energy $\hbar\omega_{3\downarrow}$ due
to the magnetic field splitting in bare atomic hyperfine levels. The
Hamiltonian for rf-coupling may be written as, 
\begin{equation}
{\cal V}_{rf}=V_{0}\int d{\bf r}\left[\psi_{3}^{\dagger}\left({\bf r}\right)\psi_{\downarrow}\left({\bf r}\right)+\psi_{\downarrow}^{\dagger}\left({\bf r}\right)\psi_{3}\left({\bf r}\right)\right],
\end{equation}
 where $\psi_{3}^{\dagger}\left({\bf r}\right)$ ($\psi_{\downarrow}^{\dagger}\left({\bf r}\right)$)
is the field operator which creates an atom in $\left|3\right\rangle $
($\left|\downarrow\right\rangle $) at the position ${\bf r}$ and,
$V_{0}$ is the strength of the rf drive and is related to the Rabi
frequency $\omega_{R}$ with $V_{0}=\hbar\omega_{R}/2$.

For the rf-spectroscopy of weakly bound molecules that is of interest
in this work, a molecule is initially at rest in the bound state $\left|\Phi_{2B}\right\rangle $
with energy $E_{0}=-\epsilon_{B}$. Here $\epsilon_{B}$ is the binding
energy of the molecules. A radio-frequency photon with energy $\hbar\omega$
will break the molecule and transfer one of the atoms to the third
state $\left|3\right\rangle $. In the case that there is no interaction
between the state $\left|3\right\rangle $ and the spin-up and spin-down
states, the final state $\left|\Phi_{f}\right\rangle $ involves a
free atom in the third state and a remaining atom in the system. According
to the Fermi's golden rule, the rf strength of breaking molecules
and transferring atoms is proportional to the Franck-Condon factor
\cite{Chin2005}, 
\begin{equation}
F\left(\omega\right)=\left|\left\langle \Phi_{f}\right|{\cal V}_{rf}\left|\Phi_{2B}\right\rangle \right|^{2}\delta\left[\omega-\omega_{3\downarrow}-\frac{E_{f}-E_{0}}{\hbar}\right],\label{FC}
\end{equation}
 where the Dirac delta function ensures energy conservation and $E_{f}$
is the energy of the final state. The integrated Franck-Condon factor
over frequency should be unity, $\int_{-\infty}^{+\infty}F\left(\omega\right)d\omega=1$,
if we can find a complete set of final states for rf transition. Hereafter,
without any confusion we shall ignore the energy splitting in the
bare atomic hyperfine levels and set $\omega_{3\downarrow}=0$. To
calculate the Franck-Condon factor Eq. (\ref{FC}), it is crucial
to understand the initial two-particle bound state $\left|\Phi_{2B}\right\rangle $
and the final two-particle state $\left|\Phi_{f}\right\rangle $.

\section{Equal Rashba and Dresselhaus spin-orbit coupling}

Let us first consider a spin-orbit-coupled atomic Fermi gas realized
recently at Shanxi University \cite{exptShanXi} and at MIT \cite{exptMIT}.
In these two experiments, the spin-orbit coupling is induced by the
spatial dependence of two counter propagating Raman laser beams that
couple the two spin states of the system. Near Feshbach resonances,
the system may be described by a model Hamiltonian ${\cal H}={\cal H}_{0}+{\cal H}_{int}$,
where 
\begin{eqnarray}
{\cal H}_{0} & = & \sum_{\sigma}\int d{\bf r}\,\psi_{\sigma}^{\dagger}\left({\bf r}\right)\frac{\hbar^{2}\hat{\mathbf{k}}^{2}}{2m}\psi_{\sigma}\left({\bf r}\right)+\nonumber \\
 &  & \int d{\bf r}\left[\psi_{\uparrow}^{\dagger}\left({\bf r}\right)\left(\frac{\Omega_{R}}{2}e^{i2k_{R}x}\right)\psi_{\downarrow}\left({\bf r}\right)+\text{H.c.}\right]\label{bareHami1}
\end{eqnarray}
 is the single-particle Hamiltonian and 
\begin{equation}
{\cal H}_{int}=U_{0}\int d{\bf r}\psi_{\uparrow}^{\dagger}\left({\bf r}\right)\psi_{\downarrow}^{\dagger}\left({\bf r}\right)\psi_{\downarrow}\left({\bf r}\right)\psi_{\uparrow}\left({\bf r}\right)
\end{equation}
 is the interaction Hamiltonian describing the contact interaction
between two spin states. Here, $\psi_{\sigma}^{\dagger}\left({\bf r}\right)$
is the creation field operator for atoms in the spin-state $\sigma$,
$\hbar\mathbf{\hat{k}}\equiv-i\hbar\mathbf{\nabla}$ is the momentum
operator, $\Omega_{R}$ is the coupling strength of Raman beams, $k_{R}$
$=2\pi/\lambda_{R}$ is determined by the wave length $\lambda_{R}$
of two Raman lasers and therefore $2\hbar k_{R}$ is the momentum
transfer during the two-photon Raman process. The interaction strength
is denoted by the bare interaction parameter $U_{0}$. It should be
regularized in terms of the \textit{s}-wave scattering length $a_{s}$,
i.e., $1/U_{0}=m/(4\pi\hbar^{2}a_{s})-\sum_{{\bf k}}m/(\hbar^{2}\mathbf{k}^{2})$.

To solve the many-body Hamiltonian Eq. (\ref{bareHami1}), it is useful
to remove the spatial dependence of the Raman coupling term, by introducing
the following new field operators $\tilde{\psi}_{\sigma}$ via 
\begin{eqnarray}
\psi_{\uparrow}\left({\bf r}\right) & = & e^{+ik_{R}x}\tilde{\psi}_{\uparrow}\left({\bf r}\right),\label{eq:gauge1}\\
\psi_{\downarrow}\left({\bf r}\right) & = & e^{-ik_{R}x}\tilde{\psi}_{\downarrow}\left({\bf r}\right).\label{eq:gauge2}
\end{eqnarray}
 With the new field operators $\tilde{\psi}_{\sigma}$, the single-particle
Hamiltonian then becomes, 
\begin{eqnarray}
{\cal H}_{0} & = & \sum_{\sigma}\int d{\bf r}\tilde{\psi}_{\sigma}^{\dagger}\left({\bf r}\right)\frac{\hbar^{2}\left({\bf \hat{k}}\pm k_{R}{\bf e}_{x}\right)^{2}}{2m}\tilde{\psi}_{\sigma}\left({\bf r}\right)\nonumber \\
 &  & +\frac{\Omega_{R}}{2}\int d{\bf r}\left[\tilde{\psi}_{\uparrow}^{\dagger}\left({\bf r}\right)\tilde{\psi}_{\downarrow}\left({\bf r}\right)+\text{H.c.}\right],\label{bareHami2}
\end{eqnarray}
 where in the first term on the right hand side of the equation we
take ``$+$'' for spin-up atoms and ``$-$'' for spin-down atoms.
The form of the interaction Hamiltonian is invariant, 
\begin{equation}
{\cal H}_{int}=U_{0}\int d{\bf r}\,\tilde{\psi}_{\uparrow}^{\dagger}\left({\bf r}\right)\tilde{\psi}_{\downarrow}^{\dagger}\left({\bf r}\right)\tilde{\psi}_{\downarrow}\left({\bf r}\right)\tilde{\psi}_{\uparrow}\left({\bf r}\right).\label{intHami}
\end{equation}
 However, the rf Hamiltonian acquires an effective momentum transfer
$k_{R}{\bf e}_{x}$, 
\begin{equation}
{\cal V}_{rf}=V_{0}\int d{\bf r}\left[e^{-ik_{R}x}\psi_{3}^{\dagger}\left({\bf r}\right)\tilde{\psi}_{\downarrow}\left({\bf r}\right)+\text{H.c.}\right].
\end{equation}
 For later reference, we shall rewrite the rf Hamiltonian in terms
of field operators in the momentum space, 
\begin{equation}
{\cal V}_{rf}=V_{0}\sum_{{\bf q}}\left(c_{{\bf q}-k_{R}{\bf e}_{x},3}^{\dagger}c_{{\bf q}\downarrow}+\text{H.c.}\right),
\end{equation}
 where $\psi_{3}^{\dagger}\left({\bf r}\right)\equiv\sum_{{\bf q}}c_{{\bf q}3}^{\dagger}e^{i{\bf q}\cdot{\bf r}}$
and $\tilde{\psi}_{\downarrow}\left({\bf r}\right)\equiv\sum_{{\bf q}}c_{{\bf q}\downarrow}e^{i{\bf q}\cdot{\bf r}}$.
Hereafter, we shall denote $c_{{\bf k}3}$ and $c_{{\bf k}\sigma}$
as the field operators (in the momentum space) for atoms in the third
state and in the spin-state $\sigma$, respectively.

\subsection{Single-particle solution}

Using the Pauli matrices, the single-particle Hamiltonian takes the
form,\begin{widetext} 
\begin{equation}
{\cal H}_{0}=\int d{\bf r}[\tilde{\psi}_{\uparrow}^{\dagger}\left({\bf r}\right),\tilde{\psi}_{\downarrow}^{\dagger}\left({\bf r}\right)]\left[\frac{\hbar^{2}\left(k_{R}^{2}+k^{2}\right)}{2m}+h\sigma_{x}+\lambda k_{x}\sigma_{z}\right]\left[\begin{array}{c}
\tilde{\psi}{}_{\uparrow}\left({\bf r}\right)\\
\tilde{\psi}_{\downarrow}\left({\bf r}\right)
\end{array}\right],\label{spHami}
\end{equation}
 where for convenience we have defined the spin-orbit coupling constant
denoted as $\lambda\equiv\hbar^{2}k_{R}/m$ and an ``effective''
Zeeman field $h\equiv\Omega_{R}/2$. This Hamiltonian is equivalent
to the one with equal Rashba and Dresselhaus spin-orbit coupling,
$\lambda k_{x}\sigma_{y}$. To see this, let us take the second transformation
and introduce new field operators $\Psi_{\sigma}\left({\bf r}\right)$
via 
\begin{eqnarray}
\tilde{\psi}_{\uparrow}\left({\bf r}\right) & = & \frac{1}{\sqrt{2}}\left[\Psi_{\uparrow}\left({\bf r}\right)-i\Psi_{\downarrow}\left({\bf r}\right)\right],\\
\tilde{\psi}_{\downarrow}\left({\bf r}\right) & = & \frac{1}{\sqrt{2}}\left[\Psi_{\uparrow}\left({\bf r}\right)+i\Psi_{\downarrow}\left({\bf r}\right)\right].
\end{eqnarray}
 Using these new field operators, the single-particle Hamiltonian
now takes the form, 
\begin{equation}
{\cal H}_{0}=\int d{\bf r[\Psi_{\uparrow}^{\dagger}\left({\bf r}\right),\Psi_{\downarrow}^{\dagger}\left({\bf r}\right)]}\left[\frac{\hbar^{2}\left(k_{R}^{2}+k^{2}\right)}{2m}+\lambda k_{x}\sigma_{y}+h\sigma_{z}\right]\left[\begin{array}{c}
\Psi_{\uparrow}\left({\bf r}\right)\\
\Psi_{\downarrow}\left({\bf r}\right)
\end{array}\right],
\end{equation}
 \end{widetext} which is precisely the Hamiltonian with equal Rashba
and Dresselhaus spin-orbit coupling.

The single-particle Hamiltonian Eq. (\ref{spHami}) can be diagonalized
to yield two eigenvalues 
\begin{equation}
E_{{\bf k\pm}}=\frac{\hbar^{2}k_{R}^{2}}{2m}+\frac{\hbar^{2}k^{2}}{2m}{\bf \pm}\sqrt{h^{2}+\lambda^{2}k_{x}^{2}}.
\end{equation}
 Here ``${\bf \pm}$'' stands for the two helicity branches. The
single-particle eigenstates or the field operators in the helicity
basis take the form 
\begin{eqnarray}
c_{{\bf k}+} & = & +c_{{\bf k}\uparrow}\cos\theta_{{\bf k}}+c_{{\bf k}\downarrow}\sin\theta_{{\bf k}},\label{ansatzRD1}\\
c_{{\bf k}-} & = & -c_{{\bf k}\uparrow}\sin\theta_{{\bf k}}+c_{{\bf k}\downarrow}\cos\theta_{{\bf k}},\label{ansatzRD2}
\end{eqnarray}
 where 
\begin{equation}
\theta_{{\bf k}}=\arctan[(\sqrt{h^{2}+\lambda^{2}k_{x}^{2}}-\lambda k_{x})/h]>0
\end{equation}
 is an angle determined by $h$ and $k_{x}$. Note that, 
\begin{eqnarray}
\cos^{2}\theta_{{\bf k}} & = & \frac{1}{2}\left(1+\frac{\lambda k_{x}}{\sqrt{h^{2}+\lambda^{2}k_{x}^{2}}}\right),\\
\sin^{2}\theta_{{\bf k}} & = & \frac{1}{2}\left(1-\frac{\lambda k_{x}}{\sqrt{h^{2}+\lambda^{2}k_{x}^{2}}}\right).
\end{eqnarray}
 Note also that the minimum energy of the single-particle energy dispersion
is given by \cite{Jiang2011} 
\begin{equation}
E_{\min}=\frac{\hbar^{2}k_{R}^{2}}{2m}-\frac{m\lambda^{2}}{2\hbar^{2}}-\frac{\hbar^{2}h^{2}}{2m\lambda^{2}}=-\frac{\hbar^{2}h^{2}}{2m\lambda^{2}},
\end{equation}
 if $h<m\lambda^{2}/\hbar^{2}$.

\subsection{The initial two-particle bound state $\left|\Phi_{2B}\right\rangle $}

In the presence of spin-orbit coupling, the wave function of the initial
two-body bound state has both spin singlet and triplet components
\cite{Vyasanakere2011,Yu2011,Hu2011}. The wave function at zero center-of-mass
momentum, $\left|\Phi_{2B}\right\rangle $, may be written as \cite{Yu2011},\begin{widetext}
\begin{equation}
\left|\Phi_{2B}\right\rangle =\frac{1}{\sqrt{2{\cal C}}}\sum_{{\bf k}}\left[\psi_{\uparrow\downarrow}\left({\bf k}\right)c_{{\bf k}\uparrow}^{\dagger}c_{-{\bf k}\downarrow}^{\dagger}+\psi_{\downarrow\uparrow}\left({\bf k}\right)c_{{\bf k}\downarrow}^{\dagger}c_{-{\bf k}\uparrow}^{\dagger}+\psi_{\uparrow\uparrow}\left({\bf k}\right)c_{{\bf k}\uparrow}^{\dagger}c_{-{\bf k}\uparrow}^{\dagger}+\psi_{\downarrow\downarrow}\left({\bf k}\right)c_{{\bf k}\downarrow}^{\dagger}c_{-{\bf k}\downarrow}^{\dagger}\right]\left|\text{vac}\right\rangle ,\label{PhiB}
\end{equation}
 where $c_{{\bf k}\uparrow}^{\dagger}$ and $c_{{\bf k}\downarrow}^{\dagger}$
are creation field operators of spin-up and spin-down atoms with momentum
${\bf k}$ and ${\cal C}\equiv\sum_{{\bf k}}[\left|\psi_{\uparrow\downarrow}\left({\bf k}\right)\right|^{2}+\left|\psi_{\downarrow\uparrow}\left({\bf k}\right)\right|^{2}+\left|\psi_{\uparrow\uparrow}\left({\bf k}\right)\right|^{2}+\left|\psi_{\downarrow\downarrow}\left({\bf k}\right)\right|^{2}]$
is the normalization factor. From the Schrödinger equation $({\cal H}_{0}+{\cal H}_{int})\,\left|\Phi_{2B}\right\rangle =E_{0}\,\left|\Phi_{2B}\right\rangle $,
we can straightforwardly derive the following equations for coefficients
$\psi_{\sigma\sigma'}$ appearing in the above two-body wave function
\cite{Yu2011}: 
\begin{eqnarray}
\left[E_{0}-\left(\frac{\hbar^{2}k_{R}^{2}}{m}+\frac{\hbar^{2}k^{2}}{m}+2\lambda k_{x}\right)\right]\psi_{\uparrow\downarrow}\left({\bf k}\right) & = & +\frac{U_{0}}{2}\sum\limits _{{\bf k}^{\prime}}\left[\psi_{\uparrow\downarrow}\left({\bf k}^{\prime}\right)-\psi_{\downarrow\uparrow}\left({\bf k}^{\prime}\right)\right]+h\psi_{\uparrow\uparrow}\left({\bf k}\right)+h\psi_{\downarrow\downarrow}\left({\bf k}\right),\\
\left[E_{0}-\left(\frac{\hbar^{2}k_{R}^{2}}{m}+\frac{\hbar^{2}k^{2}}{m}-2\lambda k_{x}\right)\right]\psi_{\downarrow\uparrow}\left({\bf k}\right) & = & -\frac{U_{0}}{2}\sum\limits _{{\bf k}^{\prime}}\left[\psi_{\uparrow\downarrow}\left({\bf k}^{\prime}\right)-\psi_{\downarrow\uparrow}\left({\bf k}^{\prime}\right)\right]+h\psi_{\uparrow\uparrow}\left({\bf k}\right)+h\psi_{\downarrow\downarrow}\left({\bf k}\right),\\
\left[E_{0}-\left(\frac{\hbar^{2}k_{R}^{2}}{m}+\frac{\hbar^{2}k^{2}}{m}\right)\right]\psi_{\uparrow\uparrow}\left({\bf k}\right) & = & h\psi_{\uparrow\downarrow}\left({\bf k}\right)+h\psi_{\downarrow\uparrow}\left({\bf k}\right),\\
\left[E_{0}-\left(\frac{\hbar^{2}k_{R}^{2}}{m}+\frac{\hbar^{2}k^{2}}{m}\right)\right]\psi_{\downarrow\downarrow}\left({\bf k}\right) & = & h\psi_{\uparrow\downarrow}\left({\bf k}\right)+h\psi_{\downarrow\uparrow}\left({\bf k}\right),
\end{eqnarray}
 \end{widetext} where $E_{0}=-\epsilon_{B}<0$ is the energy of the
two-body bound state. Let us introduce $A_{{\bf k}}\equiv-\epsilon_{B}-(\hbar^{2}k_{R}^{2}/m+\hbar^{2}k^{2}/m)<0$
and different spin components of the wavefunctions, 
\begin{eqnarray}
\psi_{s}\left({\bf k}\right) & = & \frac{1}{\sqrt{2}}\left[\psi_{\uparrow\downarrow}\left({\bf k}\right)-\psi_{\downarrow\uparrow}\left({\bf k}\right)\right],\label{psis}\\
\psi_{a}\left({\bf k}\right) & = & \frac{1}{\sqrt{2}}\left[\psi_{\uparrow\downarrow}\left({\bf k}\right)+\psi_{\downarrow\uparrow}\left({\bf k}\right)\right].\label{psia}
\end{eqnarray}
 It is easy to see that, 
\begin{eqnarray}
\psi_{\uparrow\uparrow}\left({\bf k}\right) & = & \frac{\sqrt{2}h}{A_{{\bf k}}}\psi_{a}\left({\bf k}\right),\\
\psi_{\downarrow\downarrow}\left({\bf k}\right) & = & \frac{\sqrt{2}h}{A_{{\bf k}}}\psi_{a}\left({\bf k}\right),\\
\psi_{a}\left({\bf k}\right) & = & \lambda k_{x}\left[\frac{1}{A_{{\bf k}}-2h}+\frac{1}{A_{{\bf k}}+2h}\right]\psi_{s}\left({\bf k}\right),
\end{eqnarray}
 and 
\begin{equation}
\left[A_{{\bf k}}-\frac{4\lambda^{2}k_{x}^{2}}{A_{{\bf k}}-4h^{2}/A_{{\bf k}}}\right]\psi_{s}\left({\bf k}\right)=U_{0}\sum_{{\bf k}^{\prime}}\psi_{s}\left({\bf k}^{\prime}\right).\label{gap}
\end{equation}
 As required by the symmetry of fermionic system, the spin-singlet
wave function $\psi_{s}\left({\bf k}\right)$ is an even function
of the momentum ${\bf k}$, i.e., $\psi_{s}(-{\bf k)}=\psi_{s}({\bf k)},$
and the spin-triplet wave functions are odd functions, satisfying
$\psi_{a}\left(-{\bf k}\right)=-\psi_{a}\left({\bf k}\right)$, $\psi_{\uparrow\uparrow}\left(-{\bf k}\right)=-\psi_{\uparrow\uparrow}\left({\bf k}\right)$
and $\psi_{\downarrow\downarrow}\left(-{\bf k}\right)=-\psi_{\downarrow\downarrow}\left({\bf k}\right)$.
The un-normalized wavefunction $\psi_{s}\left({\bf k}\right)=[A_{{\bf k}}-4\lambda^{2}k_{x}^{2}/(A_{{\bf k}}-4h^{2}/A_{{\bf k}})]^{-1}$
is given by, 
\begin{equation}
\psi_{s}\left({\bf k}\right)=\frac{1}{h^{2}+\lambda^{2}k_{x}^{2}}\left[\frac{h^{2}}{A_{{\bf k}}}+\frac{\lambda^{2}k_{x}^{2}A_{{\bf k}}}{A_{{\bf k}}^{2}-4\left(h^{2}+\lambda^{2}k_{x}^{2}\right)}\right].
\end{equation}
 Using Eq. (\ref{gap}) and un-normalized wave function $\psi_{s}\left({\bf k}\right)$,
the bound-state energy $E_{0}$ or the binding energy $\epsilon_{B}$
is determined by $U_{0}\sum_{{\bf k}}\psi_{s}({\bf k)}=1$, or more
explicitly, 
\begin{equation}
\frac{m}{4\pi\hbar^{2}a_{s}}-\sum_{{\bf k}}\left[\psi_{s}\left({\bf k}\right)+\frac{m}{\hbar^{2}k^{2}}\right]=0.
\end{equation}
 Here we have replaced the bare interaction strength $U_{0}$ by the
\textit{s}-wave scattering length $a_{s}$ using the standard regularization
scheme mentioned earlier. The normalization factor of the total two-body
wave function is given by, 
\begin{equation}
{\cal C}=\sum_{{\bf k}}\left|\psi_{s}\left({\bf k}\right)\right|^{2}\left[1+\frac{2\lambda^{2}k_{x}^{2}}{\left(A_{{\bf k}}-2h\right)^{2}}+\frac{2\lambda^{2}k_{x}^{2}}{\left(A_{{\bf k}}+2h\right)^{2}}\right].
\end{equation}

\subsection{The final two-particle state $\left|\Phi_{f}\right\rangle $}

Let us consider now the final state $\left|\Phi_{f}\right\rangle $.
For this purpose, it is useful to calculate 
\begin{equation}
{\cal V}_{rf}\left|\Phi_{2B}\right\rangle =V_{0}\sum_{{\bf q}}c_{-{\bf q}-k_{R}{\bf e}_{x},3}^{+}c_{-{\bf q}\downarrow}\left|\Phi_{2B}\right\rangle 
\end{equation}
 and then determine possible final states. It can be readily seen
that, \begin{widetext} 
\begin{equation}
{\cal V}_{rf}\left|\Phi_{2B}\right\rangle =-\frac{V_{0}}{\sqrt{2{\cal C}}}\sum_{{\bf q}}c_{-{\bf q}-k_{R}{\bf e}_{x},3}^{+}\left\{ \left[\psi_{\uparrow\downarrow}\left({\bf q}\right)-\psi_{\downarrow\uparrow}\left(-{\bf q}\right)\right]c_{{\bf q}\uparrow}^{\dagger}+\left[\psi_{\downarrow\downarrow}\left({\bf q}\right)-\psi_{\downarrow\downarrow}\left(-{\bf q}\right)\right]c_{{\bf q}\downarrow}^{\dagger}\right\} \left|\text{vac}\right\rangle .
\end{equation}
 Rewriting $\psi_{\uparrow\downarrow}$\textbf{\ }and $\psi_{\downarrow\uparrow}$
in terms of $\psi_{s}$ and $\psi_{a}$ as shown in Eqs.~(\ref{psis})
and (\ref{psia}), and exploiting the parity of the wave functions,
we obtain a general result valid for any type of spin-orbit coupling,
\begin{equation}
{\cal V}_{rf}\left|\Phi_{2B}\right\rangle =-\sqrt{\frac{1}{{\cal C}}}V_{0}\sum_{{\bf q}}c_{-{\bf q}-k_{R}{\bf e}_{x},3}^{+}\left\{ \left[\psi_{s}\left({\bf q}\right)+\psi_{a}\left({\bf q}\right)\right]c_{{\bf q}\uparrow}^{\dagger}+\sqrt{2}\psi_{\downarrow\downarrow}\left({\bf q}\right)c_{{\bf q}\downarrow}^{\dagger}\right\} \left|\text{vac}\right\rangle .
\end{equation}
 To proceed, we need to rewrite the field operators $c_{{\bf q}\uparrow}^{\dagger}$
and $c_{{\bf q}\downarrow}^{\dagger}$ in terms of creation operators
in the helicity basis. For the case of equal Rashba and Dresselhaus
spin-orbit coupling, using Eqs. (\ref{ansatzRD1}) and (\ref{ansatzRD2}),
we find that 
\begin{eqnarray}
c_{{\bf q}\uparrow}^{\dagger} & = & \cos\theta_{{\bf q}}c_{{\bf q}+}^{\dagger}-\sin\theta_{{\bf q}}c_{{\bf q}-}^{\dagger},\\
c_{{\bf q}\downarrow}^{\dagger} & = & \sin\theta_{{\bf q}}c_{{\bf q}+}^{\dagger}+\cos\theta_{{\bf q}}c_{{\bf q}-}^{\dagger}.
\end{eqnarray}
 Thus, we obtain 
\begin{equation}
{\cal V}_{rf}\left|\Phi_{2B}\right\rangle =-\sqrt{\frac{1}{{\cal C}}}V_{0}\sum_{{\bf q}}c_{-{\bf q}-k_{R}{\bf e}_{x},3}^{+}\left[s_{{\bf q}+}c_{{\bf q}+}^{\dagger}-s_{{\bf q}-}c_{{\bf q}-}^{\dagger}\right]\left|\text{vac}\right\rangle ,\label{finalstateRD}
\end{equation}
 where 
\begin{eqnarray}
s_{{\bf q}+} & = & \left[\psi_{s}\left({\bf q}\right)+\psi_{a}\left({\bf q}\right)\right]\cos\theta_{{\bf q}}+\sqrt{2}\psi_{\downarrow\downarrow}\left({\bf q}\right)\sin\theta_{{\bf q}},\\
s_{{\bf q}-} & = & \left[\psi_{s}\left({\bf q}\right)+\psi_{a}\left({\bf q}\right)\right]\sin\theta_{{\bf q}}-\sqrt{2}\psi_{\downarrow\downarrow}\left({\bf q}\right)\cos\theta_{{\bf q}}.
\end{eqnarray}
 Eq. (\ref{finalstateRD}) can be interpreted as follows. The rf photon
breaks a stationary molecule and transfers a spin-down atom to the
third state. We have two possible final states: (1) we may have two
atoms in the third state and the upper helicity state, respectively,
with a possibility of $\left|s_{{\bf q}+}\right|^{2}/{\cal C}$; and
(2) we may also have a possibility of $\left|s_{{\bf q}-}\right|^{2}/{\cal C}$
for having two atoms in the third state and the lower helicity state,
respectively.

\subsection{Momentum-resolved rf spectroscopy}

Taking into account these two final states and using the Fermi's golden
rule, we end up with the following expression for the Franck-Condon
factor, 
\begin{equation}
F\left(\omega\right)=\frac{1}{{\cal C}}\sum_{{\bf q}}\left[s_{{\bf q}+}^{2}\delta\left(\omega-\frac{{\cal E}_{{\bf q}+}}{\hbar}\right)+s_{{\bf q}-}^{2}\delta\left(\omega-\frac{{\cal E}_{{\bf q}-}}{\hbar}\right)\right],\label{FCRD}
\end{equation}
 where 
\begin{equation}
{\cal E}_{{\bf q}\pm}\equiv\epsilon_{B}+\frac{\hbar^{2}\left(k_{R}^{2}+q^{2}\right)}{2m}\pm\sqrt{h^{2}+\lambda^{2}q_{x}^{2}}+\frac{\hbar^{2}\left({\bf q}+k_{R}{\bf e}_{x}\right)^{2}}{2m}.
\end{equation}
 The two Dirac delta functions in Eq. (\ref{FCRD}) are due to energy
conservation. For example, the energy of the initial state (of the
stationary molecule) is $E_{0}=-\epsilon_{B}$, while the energy of
the final state is $\hbar^{2}({\bf q}+k_{R}{\bf e}_{x})^{2}/(2m)$
for the free atom in the third state and $\hbar^{2}(k_{R}^{2}+q^{2})/(2m)+\sqrt{h^{2}+\lambda^{2}q_{x}^{2}}$
for the remaining atom in the upper branch. Therefore, the rf energy
$\hbar\omega$ required to have such a transfer is given by ${\cal E}_{{\bf q}+}$,
as shown by the first Dirac delta function. It is easy to check that
the Franck-Condon factor is integrated to unity, $\int_{-\infty}^{+\infty}F\left(\omega\right)=1$.

Experimentally, in addition to measuring the total number of atoms
transferred to the third state, which is proportional to $F(\omega)$,
we may also resolve the transferred number of atoms for a given momentum
or wave-vector $k_{x}$. Such a momentum-resolved rf spectroscopy
has already been implemented for a non-interacting spin-orbit coupled
atomic Fermi gas at Shanxi University and at MIT. Accordingly, we
may define a momentum-resolved Franck-Condon factor, 
\begin{equation}
F\left(k_{x},\omega\right)=\frac{1}{{\cal C}}\sum_{{\bf q}_{\perp}}\left[s_{{\bf q}+}^{2}\delta\left(\omega-\frac{{\cal E}_{{\bf q}+}}{\hbar}\right)+s_{{\bf q}-}^{2}\delta\left(\omega-\frac{{\cal E}_{{\bf q}-}}{\hbar}\right)\right],
\end{equation}
 where the summation are now over the wave-vector ${\bf q}_{\perp}\equiv(q_{y},q_{z})$
and we have defined $k_{x}\equiv q_{x}+k_{R}$ by shifting the wave-vector
$q_{x}$ by an amount $k_{R}$. This shift is due to the gauge transformation
used in Eqs. (\ref{eq:gauge1}) and (\ref{eq:gauge2}). With the help
of the two Dirac delta functions, the summation over ${\bf q}_{\perp}$
may be done analytically. We finally arrive at, 
\begin{equation}
F\left(k_{x},\omega\right)=\frac{m}{8\pi^{2}\hbar{\cal C}}\left[s_{{\bf q}+}^{2}\Theta\left(q_{\perp,+}^{2}\right)+s_{{\bf q}-}^{2}\Theta\left(q_{\perp,-}^{2}\right)\right],\label{kxFCRD}
\end{equation}
 where $\Theta\left(x\right)$ is the step function and 
\begin{equation}
q_{\perp,\pm}^{2}=\frac{m}{\hbar}\left(\omega-\frac{\epsilon_{B}}{\hbar}\right)-\left(k_{R}^{2}+q_{x}^{2}+q_{x}k_{R}\pm\frac{m}{\hbar^{2}}\sqrt{h^{2}+\lambda^{2}q_{x}^{2}}\right).\label{qprepThreshold}
\end{equation}
 \end{widetext} It is understood that we will use ${\bf q=(}q_{x},q_{\perp,+}{\bf )}$
in the calculation of $s_{{\bf q}+}$ and ${\bf q=(}q_{x},q_{\perp,-}{\bf )}$
in $s_{{\bf q}-}$.

We may immediately realize from the above expression that the momentum-resolved
Franck-Condon factor is an asymmetric function of $k_{x}$, due to
the coexistence of spin-singlet and spin-triplet wave functions in
the initial two-body bound state. Moreover, the contribution from
two final states or two branches should manifest themselves in the
different frequency domain in rf spectra. As we shall see below, these
features give us clear signals of anisotropic bound molecules formed
by spin-orbit coupling. On the other hand, from Eq. (\ref{kxFCRD}),
it is readily seen that once the momentum-resolved rf spectroscopy
is measured with high resolution, it is possible to determine precisely
$s_{{\bf q}+}^{2}$ and $s_{{\bf q}-}^{2}$ and then re-construct
the two-body wave function of spin-orbit bound molecules.

\subsection{Numerical results and discussions}

For equal Rashba and Dresselhaus spin-orbit coupling, the bound molecular
state exists only on the BEC side of Feshbach resonances with a positive
\textit{s}-wave scattering length, $a_{s}>0$ \cite{Jiang2011}. Thus,
it is convenient to take the characteristic binding energy $E_{B}=\hbar^{2}/(ma_{s}^{2})$
as the unit for energy and frequency. For wave-vector, we use $k_{R}=m\lambda/\hbar^{2}$
as the unit. The strength of spin-orbit coupling may be measured by
the ratio 
\begin{equation}
\frac{E_{\lambda}}{E_{B}}=\left[\frac{\hbar^{2}}{m\lambda a_{s}}\right]^{-2},
\end{equation}
 where we have defined the characteristic spin-orbit energy $E_{\lambda}\equiv m\lambda^{2}/\hbar^{2}=\hbar^{2}k_{R}^{2}/m$.
Note that, the spin-orbit coupling is also controlled by the effective
Zeeman field $h=\Omega_{R}/2$. In particular, in the limit of zero
Zeeman field $\Omega_{R}=0$, there is no spin-orbit coupling term
as shown in the original Hamiltonian Eq. (\ref{bareHami1}), although
the characteristic spin-orbit energy $E_{\lambda}\neq0$. Using $k_{R}$
and $E_{B}$ as the units for wave-vector and energy, we can write
a set of dimensionless equations for the binding energy $\epsilon_{B}=-E_{0}$,
normalization factor ${\cal C}$, Franck-Condon factor $F\left(\omega\right)$
and the momentum-resolved Franck-Condon factor $F\left(\omega,k_{x}\right)$.
We then solve them for given parameters $E_{\lambda}/E_{B}$ and $h/E_{\lambda}$.
In accord with the normalization condition $\int_{-\infty}^{+\infty}F\left(\omega\right)=1$,
the units for $F\left(\omega\right)$ and $F\left(k_{x},\omega\right)$
are taken to be $1/E_{B}$ and $1/(E_{B}k_{R})$, respectively.

\begin{figure}[htp]
\begin{centering}
\includegraphics[clip,width=0.4\textwidth]{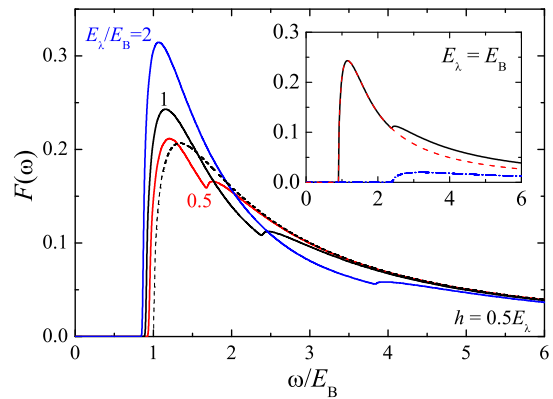} 
\par\end{centering}

\caption{(color online) Franck-Condon factor of weakly bound molecules formed
by equal Rashba and Dresselhaus spin-orbit coupling, in units of $E_{B}^{-1}$.
Here we take $h=E_{\lambda}/2$ or $\Omega_{R}=\hbar^{2}k_{R}^{2}/m$
and set $E_{\lambda}/E_{B}=0.5$, $1$, and $2$. The result without
spin-orbit coupling is plotted by the thin dashed line. The Inset
shows the different contribution from the two final states at $E_{\lambda}/E_{B}=1$.
The one with a remaining atom in the lower (upper) helicity branch
is plotted by the dashed (dot-dashed) line.}

\label{fig1} 
\end{figure}

Fig. 1 displays the Franck-Condon factor as a function of the rf frequency
at $h/E_{\lambda}=0.5$ and at several ratios of $E_{\lambda}/E_{B}$
as indicated. For comparison, we show also the rf line-shape without
spin-orbit coupling \cite{Chin2005}, $F\left(\omega\right)=(2/\pi)\sqrt{\omega-E_{B}}/\omega^{2}$,
by the thin dashed line. In the presence of spin-orbit coupling, the
existence of two possible final states is clearly revealed by the
two peaks in the rf spectra. This is highlighted in the inset for
$E_{\lambda}/E_{B}=1$, where the contribution from the two possible
final states is plotted separately. The main rf response is from the
final state with the remaining atom staying in the lower helicity
branch, i.e., the second term in the Franck-Condon factor Eq. (\ref{FCRD}).
The two peak positions may be roughly estimated from Eq. (\ref{qprepThreshold})
for the threshold frequency $\omega_{c\pm}$ of two branches, 
\begin{equation}
\hbar\omega_{c\pm}=\epsilon_{B}+\left[\frac{\hbar^{2}\left(k_{R}^{2}+q_{x}^{2}+q_{x}k_{R}\right)}{m}\pm\sqrt{h^{2}+\lambda^{2}q_{x}^{2}}\right]_{\min}.
\end{equation}
 With increasing spin-orbit coupling, the low-frequency peak becomes
more and more pronounced and shifts slightly towards lower energy.
In contrast, the high-frequency peak has a rapid blue-shift.

\begin{figure}[htp]
\begin{centering}
\includegraphics[clip,width=0.4\textwidth]{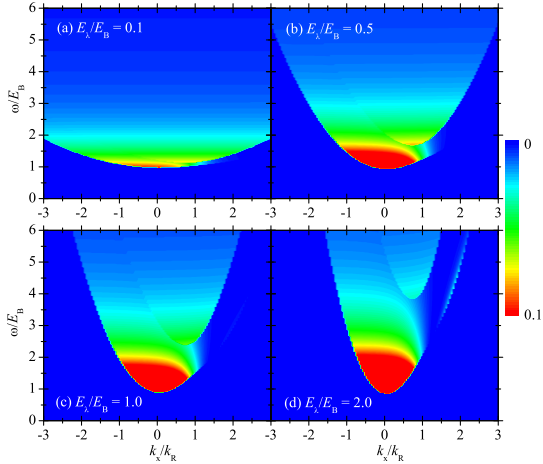} 
\par\end{centering}

\caption{(color online) Linear contour plot of momentum-resolved Franck-Condon
factor, in units of $(E_{B}k_{R})^{-1}$. Here we take $h=E_{\lambda}/2$
and consider $E_{\lambda}/E_{B}=0.1$, $0.5$, $1$, and $2$.}

\label{fig2} 
\end{figure}

\begin{figure}[htp]
\begin{centering}
\includegraphics[clip,width=0.4\textwidth]{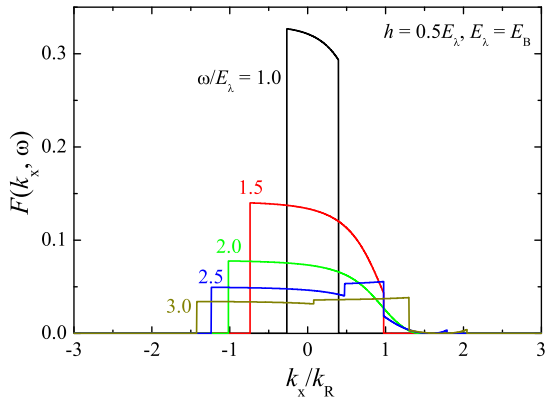} 
\par\end{centering}

\caption{(color online) Energy distribution curve of the momentum-resolved
Franck-Condon factor, in units of $(E_{B}k_{R})^{-1}$. We consider
several values of the rf frequency $\omega$ as indicated, under given
parameters $h=E_{\lambda}/2$ and $E_{\lambda}=E_{B}$.}

\label{fig3} 
\end{figure}

Fig. 2 presents the corresponding momentum-resolved Franck-Condon
factor. We find a strong asymmetric distribution as a function of
the momentum $k_{x}$. In particular, the contribution from two final
states are well separated in different frequency domains and therefore
should be easily observed experimentally. The asymmetric distribution
of $F\left(k_{x},\omega\right)$ is mostly evident in energy distribution
curve, as shown in Fig. 3, where we plot $F\left(k_{x},\omega\right)$
as a function of $k_{x}$ at several given frequencies $\omega$.
In the experiment, each of these energy distribution curves can be
obtained by a single-shot measurement.

\begin{figure}[htp]
\begin{centering}
\includegraphics[clip,width=0.4\textwidth]{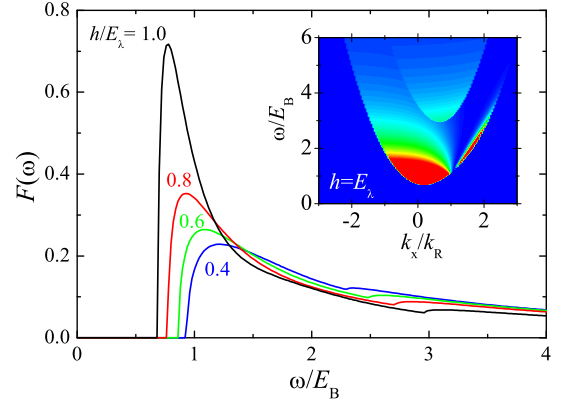} 
\par\end{centering}

\caption{(color online) Zeeman-field dependence of the Franck-Condon factor
at $E_{\lambda}=E_{B}$. Here we vary the effective Zeeman fields,
$h/E_{\lambda}=0.4$, $0.6$, $0.8$ and $1.0$. The inset shows the
momentum-resolved Franck-Condon factor at $h=E_{\lambda}$.}

\label{fig4} 
\end{figure}

We finally discuss the effect of the effective Zeeman field $h=\Omega_{R}/2$.
Fig. 4 shows how the line-shape of Franck-Condon factor evolves as
a function of the Zeeman field at $E_{\lambda}/E_{B}=1$. In general,
the larger Zeeman field the stronger spin-orbit coupling. Therefore,
as the same as in Fig. 1, the increase in Zeeman field leads to a
pronounced peak at about the binding energy. There is a red-shift
in the peak position as the binding energy becomes smaller as the
Zeeman field increases. As anticipated, the larger the Zeeman field,
the more asymmetric $F\left(k_{x},\omega\right)$ becomes. In the
inset, we show as an example the contour plot of $F\left(k_{x},\omega\right)$
at $h/E_{\lambda}=1$.

\section{Rashba spin-orbit coupling}

We now turn to the case with pure Rashba spin-orbit coupling, $\lambda(k_{y}\sigma_{x}-k_{x}\sigma_{y})$,
which may be realized experimentally in the near future. The single-particle
Hamiltonian may be written as \cite{Hu2011}, 
\begin{equation}
{\cal H}_{0}=\int d{\bf r}\left[\psi_{\uparrow}^{\dagger}\left({\bf r}\right),\psi_{\downarrow}^{\dagger}\left({\bf r}\right)\right]{\cal S}\left[\begin{array}{c}
\psi_{\uparrow}\left({\bf r}\right)\\
\psi_{\downarrow}\left({\bf r}\right)
\end{array}\right],
\end{equation}
 where the matrix 
\begin{equation}
{\cal S}=\left[\begin{array}{cc}
\hbar^{2}\left(k_{R}^{2}+k^{2}\right)/\left(2m\right) & i\lambda\left(k_{x}-ik_{y}\right)\\
-i\lambda\left(k_{x}+ik_{y}\right) & \hbar^{2}\left(k_{R}^{2}+k^{2}\right)/\left(2m\right)
\end{array}\right].
\end{equation}
 Here $\lambda$ is the coupling strength of Rashba spin-orbit coupling,
$k_{R}\equiv m\lambda/\hbar^{2}$, and we have added a constant term
$\hbar^{2}k_{R}^{2}/(2m)$ to make the minimum single-particle energy
zero \cite{Jiang2011}, i.e., $E_{\min}=0$.

\subsection{Single-particle solution}

We diagonalize the matrix ${\cal S}$ to obtain two helicity eigenvalues
\cite{Hu2011}, 
\begin{equation}
E_{{\bf k\pm}}=\frac{\hbar^{2}\left(k_{R}^{2}+k^{2}\right)}{2m}{\bf \pm}\lambda k_{\perp},
\end{equation}
 where $k_{\perp}\equiv\sqrt{k_{x}^{2}+k_{y}^{2}}$ and ``${\bf \pm}$\textbf{''}
stands for the two helicity branches. For later reference, the field
operators in the original spin basis and in the helicity basis are
related by, 
\begin{eqnarray}
c_{{\bf k}\uparrow}^{\dagger} & = & \frac{1}{\sqrt{2}}\left(c_{{\bf k}+}^{\dagger}+ie^{i\varphi_{{\bf k}}}c_{{\bf k}-}^{\dagger}\right),\label{ansatzRashba1}\\
c_{{\bf k}\downarrow}^{\dagger} & = & \frac{1}{\sqrt{2}}\left(ie^{-i\varphi_{{\bf k}}}c_{{\bf k}+}^{\dagger}+c_{{\bf k}-}^{\dagger}\right).\label{ansatzRashba2}
\end{eqnarray}
 Here $\varphi_{{\bf k}}\equiv\arg(k_{x},k_{y})$ is the azimuthal
angle of the wave-vector ${\bf k}_{\perp}$ in the $x-y$ plane.

\subsection{The initial two-particle bound state $\left|\Phi_{2B}\right\rangle $}

In the case of Rashba spin-orbit coupling, the two-body wave function
can still be written in the same form as in the previous case, i.e.,
Eq.~(\ref{PhiB}), and the Schrödinger equation leads to \cite{Yu2011},
\begin{widetext} 
\begin{eqnarray}
A_{{\bf k}}\psi_{\uparrow\downarrow}\left({\bf k}\right) & = & +\frac{U_{0}}{2}\sum_{{\bf k}^{\prime}}\left[\psi_{\uparrow\downarrow}\left({\bf k}^{\prime}\right)-\psi_{\downarrow\uparrow}\left({\bf k}^{\prime}\right)\right]-\lambda\left(k_{y}-ik_{x}\right)\psi_{\uparrow\uparrow}\left({\bf k}\right)+\lambda\left(k_{y}+ik_{x}\right)\psi_{\downarrow\downarrow}\left({\bf k}\right),\\
A_{{\bf k}}\psi_{\downarrow\uparrow}\left({\bf k}\right) & = & -\frac{U_{0}}{2}\sum_{{\bf k}^{\prime}}\left[\psi_{\uparrow\downarrow}\left({\bf k}^{\prime}\right)-\psi_{\downarrow\uparrow}\left({\bf k}^{\prime}\right)\right]+\lambda\left(k_{y}-ik_{x}\right)\psi_{\uparrow\uparrow}\left({\bf k}\right)-\lambda\left(k_{y}+ik_{x}\right)\psi_{\downarrow\downarrow}\left({\bf k}\right),\\
A_{{\bf k}}\psi_{\uparrow\uparrow}\left({\bf k}\right) & = & -\lambda\left(k_{y}+ik_{x}\right)\psi_{\uparrow\downarrow}\left({\bf k}\right)+\lambda\left(k_{y}+ik_{x}\right)\psi_{\downarrow\uparrow}\left({\bf k}\right),\\
A_{{\bf k}}\psi_{\downarrow\downarrow}\left({\bf k}\right) & = & +\lambda\left(k_{y}-ik_{x}\right)\psi_{\uparrow\downarrow}\left({\bf k}\right)-\lambda\left(k_{y}-ik_{x}\right)\psi_{\downarrow\uparrow}\left({\bf k}\right),
\end{eqnarray}
 \end{widetext} where $A_{{\bf k}}\equiv E_{0}-(\hbar^{2}k_{R}^{2}/m+\hbar^{2}k^{2}/m)<0$.
It is easy to show that $\psi_{a}\left({\bf k}\right)=0$ and 
\begin{equation}
\left[A_{{\bf k}}-\frac{4\lambda^{2}k_{\perp}^{2}}{A_{{\bf k}}}\right]\psi_{s}\left({\bf k}\right)=U_{0}\sum_{{\bf k}^{\prime}}\psi_{s}\left({\bf k}^{\prime}\right).
\end{equation}
 Thus, we obtain the (un-normalized) wavefunction: 
\begin{equation}
\psi_{s}\left({\bf k}\right)=\frac{1}{2}\left[\frac{1}{E_{0}-2E_{{\bf k+}}}+\frac{1}{E_{0}-2E_{{\bf k-}}}\right],
\end{equation}
 and the equation for the energy $E_{0}$, 
\begin{equation}
\frac{m}{4\pi\hbar^{2}a_{s}}=\sum_{{\bf k}}\left[\frac{1/2}{E_{0}-2E_{{\bf k+}}}+\frac{1/2}{E_{0}-2E_{{\bf k-}}}+\frac{m}{\hbar^{2}k^{2}}\right].
\end{equation}
 The spin-triplet wave functions $\psi_{\uparrow\uparrow}\left({\bf k}\right)$
and $\psi_{\downarrow\downarrow}\left({\bf k}\right)$ are given by,
\begin{eqnarray}
\psi_{\uparrow\uparrow}\left({\bf k}\right) & = & \left[-ie^{-i\varphi_{{\bf k}}}\frac{\sqrt{2}\lambda k_{\perp}}{E_{0}-2\epsilon_{{\bf k}}}\right]\psi_{s}\left({\bf k}\right),\\
\psi_{\downarrow\downarrow}\left({\bf k}\right) & = & \left[-ie^{+i\varphi_{{\bf k}}}\frac{\sqrt{2}\lambda k_{\perp}}{E_{0}-2\epsilon_{{\bf k}}}\right]\psi_{s}\left({\bf k}\right),
\end{eqnarray}
 where $\epsilon_{{\bf k}}\equiv\hbar^{2}k^{2}/(2m)$. The normalization
factor for the two-body wave function is therefore, 
\begin{equation}
{\cal C}=\sum_{{\bf k}}\left|\psi_{s}\left({\bf k}\right)\right|^{2}\left[1+\frac{4\lambda^{2}k_{\perp}^{2}}{\left(E_{0}-2\epsilon_{{\bf k}}\right)^{2}}\right].
\end{equation}

\subsection{The final two-particle state $\left|\Phi_{f}\right\rangle $}

To obtain the final state, we consider again ${\cal V}_{rf}\left|\Phi_{2B}\right\rangle $.
In the present case, we assume that the rf Hamiltonian is given by,
\begin{equation}
{\cal V}_{rf}=V_{0}\sum_{{\bf q}}\left(c_{{\bf q}3}^{\dagger}c_{{\bf q}\downarrow}+c_{{\bf q}\downarrow}^{\dagger}c_{{\bf q}3}\right).
\end{equation}
 Following the same procedure as in the case of equal Rashba and Dresselhaus
coupling, it is straightforward to show that, \begin{widetext} 
\begin{equation}
{\cal V}_{rf}\left|\Phi_{2B}\right\rangle =-\sqrt{\frac{1}{{\cal C}}}V_{0}\sum_{{\bf q}}c_{-{\bf q}3}^{+}\left[\psi_{s}\left({\bf q}\right)c_{{\bf q}\uparrow}^{\dagger}+\sqrt{2}\psi_{\downarrow\downarrow}\left({\bf q}\right)c_{{\bf q}\downarrow}^{\dagger}\right]\left|\text{vac}\right\rangle .
\end{equation}
 Using Eqs. (\ref{ansatzRashba1}) and (\ref{ansatzRashba2}) to rewrite
$c_{{\bf q}\uparrow}^{\dagger}$ and $c_{{\bf q}\downarrow}^{\dagger}$
in terms of $c_{{\bf q}+}^{\dagger}$ and $c_{{\bf q}-}^{\dagger}$,
we obtain, 
\begin{equation}
{\cal V}_{rf}\left|\Phi_{2B}\right\rangle =-\sqrt{\frac{1}{2{\cal C}}}V_{0}\sum_{{\bf q}}\left[\frac{c_{-{\bf q}3}^{+}c_{{\bf q}+}^{\dagger}}{E_{0}-2E_{{\bf q+}}}+\frac{ie^{i\varphi_{{\bf q}}}c_{-{\bf q}3}^{+}c_{{\bf q}-}^{\dagger}}{E_{0}-2E_{{\bf q-}}}\right]\left|\text{vac}\right\rangle .
\end{equation}
 Therefore, we have again two final states, differing in the helicity
branch that the remaining atom stays. The remaining atom stays in
the upper branch with probability $(2{\cal C})^{-1}(E_{0}-2E_{{\bf q+}})^{-2}$,
and in the lower branch with probability $(2{\cal C})^{-1}(E_{0}-2E_{{\bf q-}})^{-2}$.

\subsection{Momentum-resolved rf spectroscopy}

Using the Fermi's golden rule, we have immediately the Franck-Condon
factor, 
\begin{equation}
F\left(\omega\right)=\frac{1}{{\cal C}}\sum_{{\bf k}}\left[\frac{\delta\left(\omega-{\cal E}_{{\bf k}+}/\hbar\right)}{2\left(\epsilon_{B}+2E_{{\bf k+}}\right)^{2}}+\frac{\delta\left(\omega-{\cal E}_{{\bf k}-}/\hbar\right)}{2\left(\epsilon_{B}+2E_{{\bf k-}}\right)^{2}}\right],
\end{equation}
 where 
\begin{equation}
{\cal E}_{{\bf k}\pm}\equiv\epsilon_{B}+\frac{\hbar^{2}k_{R}^{2}}{2m}+\frac{\hbar^{2}k^{2}}{m}\pm\lambda k_{\perp}.
\end{equation}
 For Rashba spin-orbit coupling, it is reasonable to define the following
momentum-resolved Franck-Condon factor, 
\begin{equation}
F\left(k_{\perp},\omega\right)=\frac{1}{{\cal C}}\sum_{k_{z}}\left[\frac{\delta\left(\omega-{\cal E}_{{\bf k}+}/\hbar\right)}{2\left(\epsilon_{B}+2E_{{\bf k+}}\right)^{2}}+\frac{\delta\left(\omega-{\cal E}_{{\bf k}-}/\hbar\right)}{2\left(\epsilon_{B}+2E_{{\bf k-}}\right)^{2}}\right],
\end{equation}
 where we have summed over the momentum $k_{z}$. Integrating over
$k_{z}$ with the help of the two Dirac delta functions, we find that,
\begin{equation}
F\left(k_{\perp},\omega\right)=\frac{m}{16\pi^{3}\hbar{\cal C}}\left[\frac{\Theta\left(k_{z+}^{2}\right)}{\left(\hbar\omega+\hbar^{2}k_{R}^{2}/2m+\lambda k_{\perp}\right)^{2}k_{z+}}+\frac{\Theta\left(k_{z-}^{2}\right)}{\left(\hbar\omega+\hbar^{2}k_{R}^{2}/2m-\lambda k_{\perp}\right)^{2}k_{z-}}\right],
\end{equation}
 where 
\begin{equation}
k_{z\pm}^{2}=\frac{m}{\hbar}\left(\omega-\frac{\epsilon_{B}}{\hbar}\right)-\left(\frac{k_{R}^{2}}{2}+k_{\perp}^{2}\pm k_{R}k_{\perp}\right).
\end{equation}
 \end{widetext} It is easy to see that the threshold frequencies
for the two final states are given by, 
\begin{eqnarray}
\hbar\omega_{c+} & = & \epsilon_{B}+\frac{\hbar^{2}k_{R}^{2}}{2m},\\
\hbar\omega_{c-} & = & \epsilon_{B}+\frac{\hbar^{2}k_{R}^{2}}{4m},
\end{eqnarray}
 which differ by an amount of $\hbar^{2}k_{R}^{2}/(4m)=E_{\lambda}/4$.
Near $\omega_{c-}$, we find approximately that $F\left(\omega\right)\varpropto\Theta(\omega-\omega_{c-})/\omega^{2}$.
Thus, the lineshape near the threshold is similar to that of a two-dimensional
(2D) Ferm gas \cite{Zhang2012}. This similarity is related to the
fact that at low energy a 3D Fermi gas with Rashba spin-orbit coupling
has exactly the same density of states as a 2D Fermi gas \cite{Hu2012b}.

\subsection{Numerical results and discussions}

For the pure Rashba spin-orbit coupling, the molecular bound state
exists for arbitrary \textit{s}-wave scattering length $a_{s}$ \cite{Vyasanakere2011,Yu2011,Hu2011}.
We shall take $k_{R}=m\lambda/\hbar^{2}$ and $E_{\lambda}\equiv m\lambda^{2}/\hbar^{2}$
as the units for wave-vector and energy, respectively. With these
units, the dimensionless interaction strength is given by $\hbar^{2}/(m\lambda a_{s})$.
The spin-orbit effect should be mostly significant on the BCS side
with $\hbar^{2}/(m\lambda a_{s})<0$, where the bound state cannot
exist without spin-orbit coupling.

\begin{figure}[htp]
\begin{centering}
\includegraphics[clip,width=0.4\textwidth]{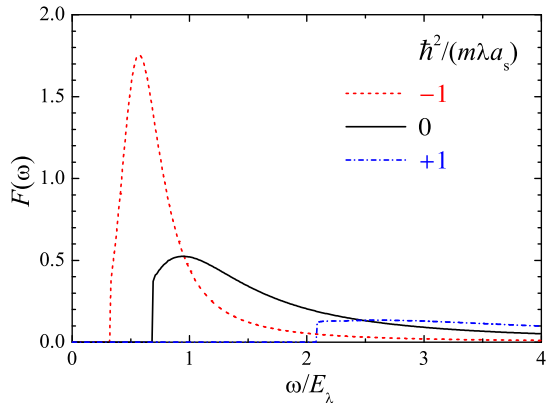} 
\par\end{centering}

\caption{(color online) Franck-Condon factor of weakly bound molecules formed
by Rashba spin-orbit coupling, in units of $E_{\lambda}^{-1}$. Here
we take $\hbar^{2}/(m\lambda a_{s})=-1$ (dashed line), $0$ (solid
line), and $1$ (dot-dashed line). In the deep BCS limit, $\hbar^{2}/(m\lambda a_{s})\rightarrow-\infty$,
the Franck-Condon factor peaks sharply at $\hbar\omega\simeq E_{\lambda}/2$
and becomes a delta-like distribution.}

\label{fig5} 
\end{figure}

Fig. 5 shows the Franck-Condon factor at three different interaction
strengths $\hbar^{2}/(m\lambda a_{s})=-1$, $0$, and $+1$. The strong
response in the BCS regime ($a_{s}<0$) or in the unitary limit ($a_{s}\rightarrow\pm\infty$)
is an unambiguous signal of the existence of Rashba molecules. In
particular, the rf line-shape in the BCS regime shows a sharp peak
at about $\hbar\omega\simeq E_{\lambda}/2$ and decays very fast at
high frequency.

\begin{figure}[htp]
\begin{centering}
\includegraphics[clip,width=0.4\textwidth]{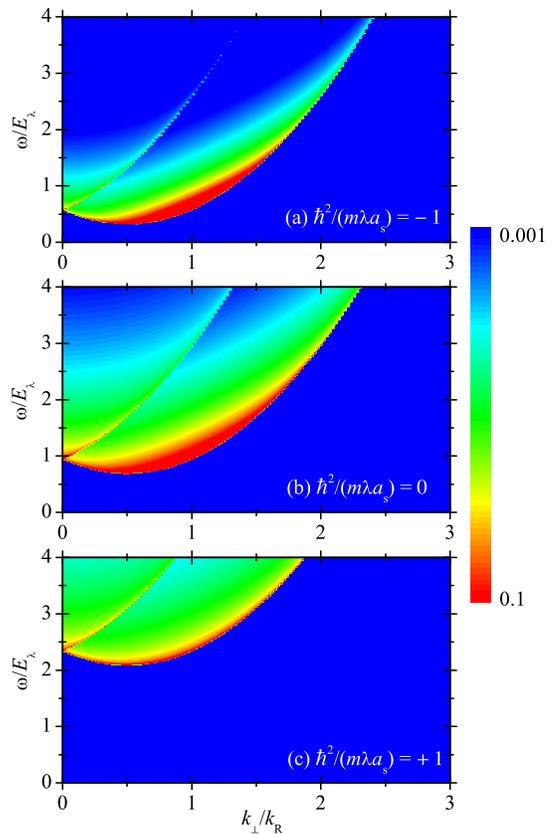} 
\par\end{centering}

\caption{(color online) Contour plot of momentum-resolved Franck-Condon factor
of weakly bound molecules formed by Rashba spin-orbit coupling, in
units of $(E_{\lambda}k_{R})^{-1}$. The intensity increases from
blue to red in a logarithmic scale. We consider $\hbar^{2}/(m\lambda a_{s})=-1$
(a), $0$ (b), and $1$ (c).}

\label{fig6} 
\end{figure}

\begin{figure}[htp]
\begin{centering}
\includegraphics[clip,width=0.4\textwidth]{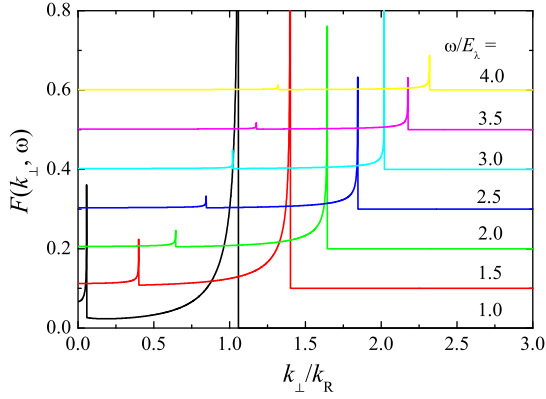} 
\par\end{centering}

\caption{(color online) Momentum-resolved Franck-Condon factor of Rashba molecules
at $\hbar^{2}/(m\lambda a_{s})=0$, shown in the form of energy distribution
curves at several rf frequencies as indicated.}

\label{fig7} 
\end{figure}

In Fig. 6, we present the corresponding momentum-resolved Franck-Condon
factor $F\left(k_{\perp},\omega\right)$, in the form of contour plots.
We can see clearly the different response from the two final states.
The momentum-resolved rf spectroscopy is particularly useful to identify
the contribution from the final state that has a remaining atom in
the upper branch, which, being integrated over $k_{\perp}$, becomes
too weak to be resolved in the total rf spectroscopy. Finally, we
report in Fig. 7 energy distribution curves of $F\left(k_{\perp},\omega\right)$
in the unitary limit $\hbar^{2}/(m\lambda a_{s})=0$. We find two
sharp peaks in each energy distribution curve, arising from the two
final states. When measured experimentally, these sharp peaks would
become much broader owing to the finite experimental energy resolution.

\section{Conclusions}

In conclusions, we have investigated theoretically the radio-frequency
spectroscopy of weakly bound molecules in a spin-orbit coupled atomic
Fermi gas. The wave function of these molecules is greatly affected
by spin-orbit coupling and has both spin-singlet and spin-triplet
components. As a result, the line-shape of the total radio-frequency
spectroscopy is qualitatively different from that of the conventional
molecules at the BEC-BCS crossover without spin-orbit coupling. In
addition, the momentum-resolved radio-frequency becomes highly asymmetric
as a function of the momentum. These features are easily observable
in current experiments with spin-orbit coupled Fermi gases of $^{40}$K
atoms and $^{6}$Li atoms. On the other hand, from the high-resolution
momentum-resolved radio-frequency, we may re-construct the two-body
wave function of the bound molecules.

We consider so far the molecular response in the radio-frequency spectroscopy.
Our results should be quantiatively reliable in the deep BEC limit
with negligible number of atoms, i.e., in the interaction parameter
regime with $1/(k_{F}a_{s})>2$. However, in a real experiment, in
order to maximize the spin-orbit effect, it is better to work closer
to Feshbach resonances, i.e., $1/(k_{F}a_{s})\sim0.5$. Under this
situation, the spin-orbit coupled Fermi gas consists of both atoms
and weakly bound molecules, which may strongly interact with each
other. Our prediction for the molecular response is still qualitatively
valid, with the understanding that there would be an additional pronounced
atomic response in the rf spectra. A more in-depth investigation of
radio-frequency spectroscopy requires complicated many-body calculations
beyond our simple two-body picture pursued in the present work.

\section*{Acknowledgments}

We would like to thank Zeng-Qiang Yu and Hui Zhai for useful discussions.
HH and XJL are supported by the ARC Discovery Projects (Grant No.
DP0984522 and No. DP0984637) and the National Basic Research Program
of China (NFRP-China, Grant No. 2011CB921502). HP is supported by
the NSF and the Welch Foundation (Grant No. C-1669). JZ is supported
by the NSFC Project for Excellent Research Team (Grant No. 61121064)
and the NFRP-China (Grant No. 2011CB921601). SGP is supported by the
NSFC-China (Grant No. 11004224) and the NFRP-China (Grant No. 2011CB921601).

\end{document}